\documentclass{elsart}
 \usepackage{graphics}
 \usepackage{graphicx}
 \usepackage{epsfig}
 \usepackage{amssymb}

\begin{document}

\begin{frontmatter}

\title{\large \boldmath \bf Search for the Lepton Flavor Violation Processes
    $J/\psi \to$ $\mu\tau$ and $e\tau$}

\begin{small}
\begin{center}

\vspace{0.2cm}

M.~Ablikim$^{1}$, J.~Z.~Bai$^{1}$, Y.~Ban$^{10}$, 
J.~G.~Bian$^{1}$, X.~Cai$^{1}$, J.~F.~Chang$^{1}$, 
H.~F.~Chen$^{16}$, H.~S.~Chen$^{1}$, H.~X.~Chen$^{1}$, 
J.~C.~Chen$^{1}$
% ...
\vspace{0.2cm} 
\\(BES Collaboration)\\ 

\vspace{0.2cm}
\label{att}
$^1$ Institute of High Energy Physics, Beijing 100039, People's Republic of China\\
$^2$ China Center for Advanced Science and Technology (CCAST), Beijing 100080, 
People's Republic of China\\
$^3$ Guangxi Normal University, Guilin 541004, People's Republic of China\\
$^4$ Henan Normal University, Xinxiang 453002, People's Republic of China\\
$^5$ Huazhong Normal University, Wuhan 430079, People's Republic of China\\
$^6$ Hunan University, Changsha 410082, People's Republic of China\\
$^7$ Liaoning University, Shenyang 110036, People's Republic of China\\
$^8$ Nanjing Normal University, Nanjing 210097, People's Republic of China\\
$^9$ Nankai University, Tianjin 300071, People's Republic of China\\
$^{10}$ Peking University, Beijing 100871, People's Republic of China\\
$^{11}$ Shandong University, Jinan 250100, People's Republic of China\\
$^{12}$ Shanghai Jiaotong University, Shanghai 200030, People's Republic of China\\
$^{13}$ Sichuan University, Chengdu 610064, People's Republic of China\\
$^{14}$ Tsinghua University, Beijing 100084, People's Republic of China\\
$^{15}$ University of Hawaii, Honolulu, Hawaii 96822\\
$^{16}$ University of Science and Technology of China, Hefei 230026, People's Republic of China\\
$^{17}$ Wuhan University, Wuhan 430072, People's Republic of China\\
$^{18}$ Zhejiang University, Hangzhou 310028, People's Republic of China\\
\vspace{0.4cm}

$^{\ast}$ Visiting professor to University of Michigan, Ann Arbor, MI 48109 USA \\
$^{\dagger}$ Current address: Purdue University, West Lafayette, Indiana 47907, USA.

\vspace{0.2cm}

\end{center}
\end{small}
\normalsize

\begin{abstract}
{The lepton flavor violation processes
$J/\psi \to \mu\tau$ and $e\tau$ are searched for using a sample of
5.8$\times 10^7$ $J/\psi$ events collected with the BESII detector.
Zero and one candidate events, consistent with the estimated background,
are observed in $J/\psi \to \mu\tau, \tau\to
e\bar\nu_e\nu_{\tau}$ and $J/\psi\to e\tau,
\tau\to\mu\bar\nu_{\mu}\nu_{\tau}$ decays, respectively.  Upper
limits on the branching ratios are determined to be
$Br(J/\psi\to\mu\tau)<2.0 \times 10^{-6}$ and $Br(J/\psi \to e\tau) <
8.3 \times10^{-6}$ at the 90$\%$ confidence level (C.L.).}

\vspace{0.2cm}
\noindent{\it PACS:} 13.25.Gv, 14.40.Gx, 13.40.Hq
\end{abstract}



\end{frontmatter}

\section{Introduction}
In the Standard Model ( SM), lepton flavor is conserved, but it is
expected to be violated in many extensions of the SM, such as grand
unified models \cite{1}, supersymmetric models \cite{2},
left-right symmetric models \cite{3}, and models where electroweak
symmetry is broken dynamically \cite{4}. Recent experimental results
from Super-Kamiokande \cite{5}, SNO \cite{6}, and KamLAND \cite{7}
indicate strongly that neutrinos have masses and can mix with each
other. Consequently, lepton flavor symmetry is a broken symmetry.
There have been many studies both experimentally and theoretically on
searching for lepton flavor violating (LFV) processes \cite{8}, mainly
from $\mu$, $\tau$ and Z decays \cite{9}. Theoretical predictions
of LFV in decays of charmonium and bottomonium systems are discussed
in Refs. \cite{10,11,12}, and the search for the $J/\psi\to e\mu$ LFV
process at BESII is presented in Ref. \cite{13}. In this paper, we
report on a search for LFV via the decays $ J/\psi \to\mu\tau,\tau\to
e\bar\nu_e\nu_{\tau}$ and $J/\psi\to
e\tau,\tau\to\mu\bar\nu_{\mu}\nu_{\tau}$ using $5.8 \times 10^7
J/\psi$ events collected with the BESII detector.

\section{BES detector}
The Beijing Spectrometer (BES) \cite{14,15} is a conventional
solenoidal magnetic detector at the Beijing Electron Positron Collider
(BEPC).  The upgraded version of the BES detector, BESII, includes a
12-layer vertex chamber (VC), surrounding the beam pipe and providing
trigger information; a fourty-layer main drift chamber (MDC), located
radially outside the VC and providing trajectory and energy loss
($dE/dx$) information for charged tracks over $85\%$ of the total
solid angle; and an array of 48 scintillation counters surrounding the
MDC to measure the time-of-flight (TOF) of charged tracks with a
resolution of $\sim 200$ ps for hadrons.  The momentum resolution of
the MDC is $\sigma _p/p = 1.78$\%$ \sqrt{1+p^2}$ ($p$ in $\hbox{\rm
  GeV}/c$), and the $dE/dx$ resolution for hadron tracks is about
$8\%$.  Radially outside the TOF system is a 12 radiation length,
lead-gas barrel shower counter (BSC).  This measures the energies of
electrons and photons with an energy resolution of
$\sigma_E/E=21\%/\sqrt{E}$ ($E$ in GeV).  Outside the solenoidal coil,
which provides a 0.4~Tesla magnetic field over the tracking volume, is
an iron flux return that is instrumented with three double layers of
counters to identify muons of momentum greater than 0.5~GeV/c. It
provides coordinate measurements with resolutions in the outermost
layer of 10 cm and 12 cm in $r\phi$ and $z$.  The solid angle
coverage of the layers is 67\%, 67\%, and 63\% of $4\pi$,
respectively.

In the analysis, a GEANT3 based Monte Carlo program (SIMBES) with detailed 
consideration of detector performance (such as dead electronic channels)
is used. The consistency between data and Monte Carlo has been
checked in many high purity physics channels, and the agreement is
reasonable.

\section{Event selection}
We require candidate events for $ J/\psi \to\mu\tau,\tau\to
e\bar\nu_e\nu_{\tau}$ and $J/\psi\to
e\tau,\tau\to\mu\bar\nu_{\mu}\nu_{\tau}$ to have two well
reconstructed and oppositely charged tracks, each of which is well
fitted to a helix originating from the interaction region of $|x|
<$0.015 m, $|y| <$0.015 m, and $|z| <$ 0.15 m and with a polar angle,
$\theta$, satisfying $|\cos\theta| <$ 0.8.  To reject cosmic rays,
the time of flight difference of the two charged tracks should be less
than 4 ns.

Isolated photons are defined as those photons having an 
energy deposit in the BSC greater than 50 MeV, an angle
with any charged track greater than $15^{\circ}$,
and an angle between the direction defined by the first layer hit in
the BSC
and the developing direction of the cluster in the $xy$-plane less than
$18^{\circ}$.  There must be no isolated photon in the selected event.

Information from the BSC, TOF, and MDC ($dE/dx$) is used to select
electrons.  Fig. 1(a) shows the ratio of the energy deposited by the
electron in the BSC to its momentum ($E/P$) for Monte Carlo simulated
events, and Fig. 1(b) shows the energy deposited by the muon in the
BSC for Monte-Carlo simulated events.  To be an electron, the charged
track should have no hits in the muon counter, and the $E/P$ ratio
should be larger than 0.7. To further distinguish the electron from
hadrons, it is required that $\wp_{dE/dx}^e > \wp_{dE/dx}^\pi$,
$\wp_{dE/dx}^e > \wp_{dE/dx}^K$ and $\wp_{TOF}^e > \wp_{TOF}^p$, where
$\wp_{dE/dx}^i$ and $\wp_{TOF}^i$ are the particle identification
confidence levels for the $dE/dx$ and TOF measurements and $i$
denotes $e, \pi, K$ or $p$.
 
\begin{figure}[htp]
\centerline{
\psfig{figure=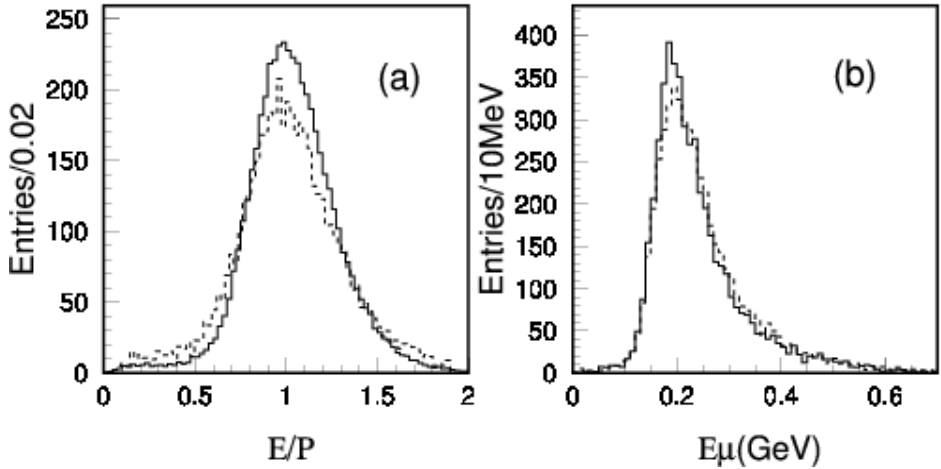,width=12.5cm,height=6.3cm,angle=0}}
\caption{(a.) Distribution of $E/P$ for electrons (MC simulation).
(b.) Distribution of energy deposited by muons in the
BSC (MC simulation). The solid histogram represents $J/\psi\to e\tau, 
\tau\to \mu\bar\nu_{\mu}\nu_{\tau}$ channel, and the dashed one is for
$J/\psi\to \mu\tau, \tau\to e\bar\nu_e\nu_{\tau}$ channel.}
\end{figure}

To select muon tracks, the differences, $\delta_i (i = r\phi, z)$,
between the closest muon hit and the projected MDC track in each layer
are used.  A good hit in the $\mu$ counter requires $|\delta_i| <
2\sigma_i$ for $i = r\phi$ and $z$. 
The total
number of good $\mu$ hits in the $\mu$ counter,
$\mu_{hit}^{good}$, can range from 0 to 3.  A track is
considered as a muon if the deposited energy in the BSC, shown in Fig.
1(b), is less than 0.3 GeV and $\mu_{hit}^{good}$ is equal to 3.

For the decay of $J/\psi \to e\tau,
\tau\to\mu\bar\nu_{\mu}\nu_{\tau}$, the momentum of the electron is
monochromatic, while that of the muon is broad, as shown in Fig. 2.
The main background for this channel comes from $J/\psi \to
(\gamma)\mu^+\mu^-$ and $e^+e^-\to(\gamma)\mu^+\mu^-$, which is shown
as the dashed histogram in Fig. 2. This background can be rejected by
requiring that the momentum of the electron $P_e$ be in the region from
1.00 to 1.08 GeV/$c$ and the momentum of the muon be less than 1.4
GeV/$c$.  Similar requirements $P_e < 1.4$ GeV/$c$ and $1.00 < P_\mu <1.08$
GeV/$c$ are applied to $J/\psi \to \mu\tau, \tau\to
e\bar\nu_e\nu_{\tau}$ candidates to suppress the background from
$J/\psi \to (\gamma)e^+e^-$ and $e^+e^-\to(\gamma)e^+e^-$. Applying
these requirements, no
candidates for $J/\psi\to\mu\tau, \tau\to e\bar\nu_e\nu_{\tau}$ and
one candidate for $J/\psi\to e\tau, \tau\to
\mu\bar\nu_{\mu}\nu_{\tau}$ survive.

\begin{figure}[htp]
\centerline{
\psfig{figure=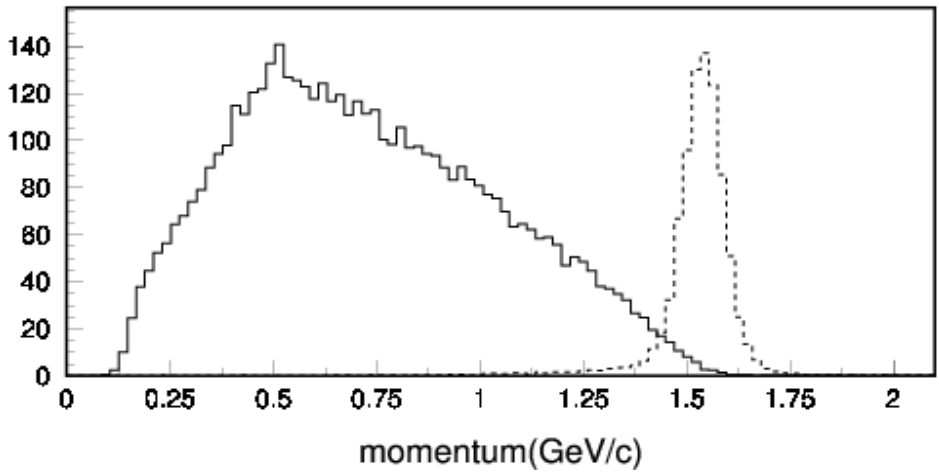,width=12.5cm,height=6.3cm,angle=0}}
\caption{Monte-Carlo distributions of muon momentum.
The solid histogram represents $J/\psi\to e\tau, 
\tau\to \mu\bar\nu_{\mu}\nu_{\tau}$ channel, and the dashed one
is for the $J/\psi\to (\gamma)\mu^+\mu^-$ channel.}
\end{figure}

\section{Efficiencies and backgrounds}

In this analysis, the $\mu$ particle identification efficiency
$\epsilon_{\mu PID}$ in the $\mu$ counter is determined using real
$\mu$ tracks. All other efficiencies, including the geometric
acceptance, momentum requirement efficiency, electron particle identification
efficiency, etc., are combined into one term, $\epsilon_{MC}$, which is
determined by Monte-Carlo simulation. The overall efficiency
is calculated as $\epsilon_{total}$=$\epsilon_{\mu PID} \times
\epsilon_{MC}$.

The $\mu$ track sample selected from $5.8 \times 10^7$ $J/\psi \to
(\gamma) \mu^+ \mu^-$ decays, as described in Ref. [13], is used
to determine the $\mu$ particle identification efficiencies
in both channels.
The $\mu$ particle identification efficiency is a function of the
transverse momentum, $P_{xy}$, of the muon.
Therefore, $\epsilon_{\mu PID}$ is determined from 
$\sum\limits_{i}\epsilon_i 
\varpi_i$, where $\epsilon_i$ is the $\mu$ particle identification  efficiency in the 
$ith$ $P_{xy}$ bin determined from the $\mu$ track sample,
and $\varpi_i$ is the weight corresponding to the number of events in
the bin determined from
the signal MC. Table 1 lists 
the $\epsilon_i$ and $\varpi_i$ in the different $P_{xy}$ regions,
and Table 2 lists the selection efficiencies.

\begin{table}[htbp]
\caption{The $\epsilon_i$ and $\varpi_i$ values in different $P_{xy}$
regions in the $J/\psi \to \mu \tau$ and e$\tau$ channels.}                                                                                
\begin{center}
\begin{tabular}{c|c|c|c} 
\hline
\hline  
  &  & {$J/\psi \to \mu \tau, \tau\to e\bar\nu_e\nu_{\tau}$}&
       {$J/\psi \to e\tau,\tau\to \mu\bar\nu_{\mu}\nu_{\tau}$} \\
\hline
 $P_{xy}$ (GeV/c)  & $\epsilon_i$ (\%) & $\varpi_i$   & $\varpi_i$ \\
\hline
$0.5 <P_{xy}<0.6$ & 0.0   &  0.0  &  13.7 \\
\hline
$0.6 <P_{xy}<0.7$ & 0.0   &  3.1  &  12.3 \\
\hline 
$0.7 <P_{xy}<0.8$ & 8.1   &  11.3 &  10.9 \\
\hline
$0.8 <P_{xy}<0.9$ & 40.6  &  17.8 &  9.7 \\
\hline
$0.9 <P_{xy}<1.0$ & 52.2  &  33.9 &  7.9 \\
\hline
$1.0 <P_{xy}<1.1$ & 53.4  &  33.9 &  6.3 \\
\hline
$1.1 <P_{xy}<1.2$ & 56.2  &  0.0  &  4.5  \\
\hline
$1.2 <P_{xy}<1.3$ & 57.7  &  0.0  &  2.8 \\
\hline
$1.3 <P_{xy}<1.4$ & 53.6  &  0.0  &  1.0 \\
\hline
\hline
\end{tabular}
\vskip 1.5cm
\end{center}
\end{table}

\begin{table}[h]
\vskip -1cm
\caption{Efficiency summary}

\begin{center}
\begin{tabular}{c|c|c} 
\hline\hline
   & J/$\psi \to \mu\tau,\tau\to e\bar\nu_e\nu_{\tau}$ (\%)
   & J/$\psi \to e\tau,\tau\to \mu\bar\nu_{\mu}\nu_{\tau}$ (\%) \\
\hline 
 $\epsilon_{\mu PID}$ & 43.9       & 17.0 \\
 $\epsilon_{MC}$     & 26.2       & 28.1 \\\hline
 $\epsilon_{total}$  & 11.5       & 4.8 \\\hline\hline
\end{tabular}
\end{center} 
\end{table}

The remaining background in both the $J/\psi \to \mu\tau,\tau\to
e\bar\nu_e\nu_{\tau}$ and $J/\psi \to e\tau,
\tau\to\mu\bar\nu_{\mu}\nu_{\tau}$ processes are studied through Monte
Carlo simulation. Almost all two-prong decay modes are generated with
5 to 10 times the number of events expected from $5.8 \times 10^7
J/\psi$ events.
For $J/\psi \to e\tau, \tau\to \mu\bar\nu_{\mu}\nu_{\tau}$, the
estimated background
is about 0.4 events from $J/\psi\to\bar{K}^*(892)^-K^+ (+ c.c.)$.
For the decay $J/\psi \to \mu\tau, \tau\to e\bar\nu_e\nu_{\tau}$,
no simulated events survive.

\section{Systematic errors}
The systematic errors in the branching ratio measurements come from
the uncertainty of the MDC tracking efficiency for charged tracks, the
error from the number of $J/\psi$ events, the differences in the
efficiencies between data and Monte-Carlo simulation for some
selection criteria, such as the electron and muon
identification criteria,
as well as
the uncertainty in $\tau$ decay branching ratio. The systematic errors
from each source are listed in Table 3; the dominant error is from muon
identification.
Adding all the systematic errors in quadrature, the total systematic errors
are 16.9$\%$ and 15.4$\%$ for $J/\psi \to
\mu\tau, \tau\to e\bar\nu_e\nu_{\tau}$ and $J/\psi\to
e\tau, \tau\to \mu\bar\nu_{\mu}\nu_{\tau}$ respectively.  

\begin{table}[htb]
\caption{ Summary of systematic errors}

\begin{center}
\begin{tabular}{ c| c c } \hline\hline
Source    & $J/\psi \to \mu\tau, \tau\to e\bar\nu_e\nu_{\tau} (\%)$ & $J/\psi \to
e\tau, \tau\to \mu\bar\nu_{\mu}\nu_{\tau} (\%$)\\ \hline 
e$_{PID}$ &3.5         & 3.3 \\ 
$\mu_{PID}$ & 15.4      & 13.7 \\
$Br(\tau$ decay) &0.3 & 0.4 \\
MDC tracking &  4.0 &4.0 \\ 
Number of $J/\psi$ events \cite{16} & 4.7 &4.7 \\ \hline 

Sum         & 16.9        &15.4   \\ \hline\hline            
\end{tabular}
\end{center}
\end{table}


\section{Results and discussion }
No $J/\psi \to\mu\tau, \tau\to e\bar\nu_e\nu_{\tau}$ candidate and one
$J/\psi\to e\tau, \tau\to\mu\bar\nu_{\mu}\nu_{\tau}$ candidate are
observed from a sample of 5.8$\times 10^7$ $J/\psi$ events, where the
estimated background events in the two chanels are 
of 0 and 0.4 events, respectively. The background events are ignored for the
conservative estimation.
 Upper limits on the branching ratios of $J/\psi \to
\mu\tau$ and $J/\psi\to e\tau$ are calculated with:
\par
$$Br(J/\psi\to X)<\frac{\lambda(N_{Signal},N_{BG})}{N_{J/\psi}\times
Br(X \to Y)\times \epsilon_{J/\psi \to X \to Y}},$$
where $X$ and $Y$ stand for the intermediate and final states, $\lambda$ is 
the upper limit on the number of observed events at the 90$\%$ C.L.,
$N_{Signal}$ and $N_{BG}$ are the numbers of observed signal and
background events respectively, $N_{J/\psi}$ represents the total
number of $J/\psi$ events, and $\epsilon$ is the detection efficiency.
The values of $\lambda(N_{Signal}$ and $N_{BG})$ can be calculated using
the method described in Refs. \cite{17} and \cite{18}.  

With the numbers summarized in Table 4, the upper 
limits on the branching ratios, after incorporating the systematic errors,
are 

\par
\begin{center}   
 $Br(J/\psi \to \mu\tau)<2.0 \times 10^{-6},$
\par
$Br(J/\psi \to e\tau)<8.3 \times 10^{-6}$
\end{center}
at the 90$\%$ C.L.
\par
Previously BES reported an upper limit on $Br(J/\psi \to e \mu)$ to be
$1.1 \times 10^{-6}$ at the 90 $\%$ C.L. \cite{13}.

\begin{table}[h]
\caption{Numbers and efficiencies used in the calculation of the upper limits.}

\begin{center}
\begin{tabular}{ c| c c } \hline\hline
   & $J/\psi \to \mu\tau, \tau\to e\bar\nu_e\nu_{\tau}$ & $J/\psi \to
 e\tau, \tau\to\mu\bar\nu_{\mu}\nu_{\tau}$\\ \hline
$N_{bg}$     &0         & 0 \\
$\epsilon (\%)$  &11.5  &4.8 \\
$N_{J/\psi}$ & $5.8\times 10^7$ & $5.8\times 10^7$ \\
$N_{Signal}$ &0         & 1 \\ 
$\lambda(N_{Signal},N_{bg})$ &2.4 & 4.0 \\
Br($\tau$ decay) ($\%$) & 17.84 & 17.37 \\\hline
Upper limit of Br. & $2.0 \times 10^{-6}$ & $8.3 \times 10^{-6}$\\\hline\hline
\end{tabular}
\end{center} 
\end{table}

In summary, the LFV processes $J/\psi \to \mu \tau$ and $e \tau$
are searched for using a sample of $5.8\times 10^7$ $J/\psi$ events. No 
candidate for $J/\psi \to \mu\tau, \tau\to e\bar\nu_e\nu_{\tau}$ and 
one candidate
for $J/\psi\to e\tau,\tau\to \mu\bar\nu_{\mu}\nu_{\tau}$,
consistent with the estimated background, are observed. 
The upper limits on the branching ratios at the 90$\%$ C.L. are
determined to be $Br(J/\psi\to\mu\tau)<2.0 \times 10^{-6}$ and
$Br(J/\psi \to e\tau) < 8.3\times 10^{-6}$.

\section{Acknowledgments}

The BES collaboration thanks the staff of BEPC and the computing 
center of IHEP for their hard efforts.
We also thank Profs. Xinmin Zhang and Jianxiong Wang for helpful
discussions. This work is supported in part by the National Natural
Science Foundation of China under contracts Nos.
19991480,10225524,
10225525, the Chinese Academy of Sciences under
contract No. KJ 95T-03, the 100 Talents Program of CAS under Contract
Nos. U-11, U-24, U-25, and the Knowledge Innovation Project of CAS
under Contract Nos. U-602, U-34 (IHEP); and by the National Natural
Science Foundation of China under Contract No.10175060(USTC), and
No.10225522(Tsinghua University).  and by the Department of Energy
under Contract No.  DE-FG03-94ER40833 (U Hawaii).

 \end{document}